# STABILITY OF INTERCELULAR EXCHANGE OF BIOCHEMICAL SUBSTANCES AFFECTED BY VARIABILITY OF ENVIRONMENTAL PARAMETERS

DRAGUTIN T. MIHAILOVIĆ, IGOR BALAZ

[1]*Faculty of Agriculture, University of Novi Sad, Dositej Obradovic Sq. 8, Novi Sad 21000, Serbia, E – mail:ibalaz@polj.uns.ac.rs, guto@polj.uns.ac.rs*

*Abstract.* Communication between cells is realized by exchange of biochemical substances. Due to internal organization of living systems and variability of external parameters, the exchange is heavily influenced by perturbations of various parameters at almost all stages of the process. Since communication is one of essential processes for functioning of living systems it is of interest to investigate conditions for its stability. Using previously developed simplified model of bacterial communication in a form of coupled difference logistic equations we investigate stability of exchange of signaling molecules under variability of internal and external parameters.

*Keywords*: Intercellular communication; substances exchange; coupled logistic equations; synchronization, robustness.

## 1. INTRODUCTION

Process of communication between cells is an excellent example of robustness in living systems. Despite heavy influence of perturbations of various internal and external parameters, it is able to maintain its functionality. Moreover it seems that living systems evolved toward ability to function undisturbed by small or moderate parameter fluctuations. It is not surprising since significant amount of fluctuations is of internal origin, due to protein disorder [1,2] and so called intrinsic noise [3,4]. Finally, due to thermal and conformational fluctuations, biochemical processes are inherently random [5]. However, it is surprising that some elaborated formal treatments of this problem are still in infancy [6,7]. It is argued here that robustness is a measure of feature persistence in systems that compels us to focus on fluctuations, and often assemblages of perturbations, qualitatively difierent in nature from those addressed by stability theory. Moreover, to address feature persistence under these sorts of perturbations, we are naturally led to study issues including: the coupling of dynamics with organizational architecture, implicit assumptions of the environment, the role of a system's evolutionary history in determining its current state and thereby its future state, the sense in which robustness characterizes the fitness of the set of "strategic options" open to the system and the capability of the system to switch among multiple functionalities [8,9]. In this paper, the following definition will be used - "robustness" is a property that allows a system to maintain its functions against internal and external perturbations. It is important to realize that robustness is concerned with maintaining *functions* of a *system* rather than *system states*, which distinguishes robustness from stability or homeostasis [7].

Previously, we developed simplified model of bacterial communication in order to investigate synchronization of substances exchange between abstract cells [10]. Since our model is inspired by a general scheme of intercellular communication, it naturally does not allow detailed modeling of some concrete, empirically verifiable intercellular communication process. Instead, it is designed to be a starting tool in a general investigation of robustness in mutually stimulative populations.

In this paper, our focus is only on question how the oscillating system which is basically stochastic, and is inherently influenced by internal and external perturbations, can maintain its functioning? In Section 2, using bacterial communication as an example, we give a short description of the model, representing cooperative communication process. In Section 3 we investigate synchronization of the model and its sensitivity to fluctuations of environmental parameters. Concluding remarks are given in Section 4.

## 2. DESCRIPTION OF THE INTERCELLULAR EXCHANGE MODEL

Signaling molecules are ones which are deliberately extracted by the cell into intracellular environment, and which can affect behavior of other cells of the same or different type (species or phenotype) by means of active uptake and subsequent changes in genetic regulations. Once appeared in intercellular environment, they can be transported to other cells that can be affected. Let us note that the term *environment*, in this paper, comprises both (i) *intracellular environment* (inside the cell) and (ii) *intercellular environment* (that surrounds cells). Since active uptake is one of the milestones of the process, a very important factor in establishing communication is a current set of receptors and transporters in cellular membrane, during the communication process. At the same time they constitute backbone of the whole process, while simultaneously are very important source of perturbations of the process due to protein disorder and intrinsic noise. Another important factor is intercellular environment which could interfere with the process of exchange. It includes: distance between cells, mechanical and dynamical properties of the fluid which serves as a channel for exchange and various abiotic and biotic factors influencing physiology of the involved cells. Finally, in order to define exchange process as communication, received molecules should induce change in genetic regulations. Therefore, concentration of signaling molecules inside of the cell, that are destined to be extracted, can serve as an indicator of dynamics of the whole process of communication. Additionally, the influence of affinity in functioning of living systems is also an important issue. In this case, it can be divided into following aspects: (a1) affinity of genetic regulators towards arriving signals which determine intensity of cellular response and (a2) affinity for uptake of signaling molecules. First aspect is genetically determined and therefore species specific.

Second aspect is more complex and is influenced by: affinity of receptors to binding specific signaling molecule, number of active receptor and their conformational fluctuations (protein disorder).

As it is obvious from the empirical description, we can infer successfulness of the communication process by monitoring: (i) number of signaling molecules, both inside and outside of the cell and (ii) their mutual influence. Concentration of signaling molecules in intercellular environment is subject to various environmental influences, and taken alone often can indicate more about state of the environment then about the communication itself. Therefore, we choose to follow concentration of signaling molecules inside of the cell as the main indicator of the process. In that case, parameters of the system are: (i) affinity $p$ by which cells perform uptake of signaling molecules (a2), that depends on number and state of appropriate receptors, (ii) concentration $c$ of signaling molecules in intercellular environment within the radius of interaction, (iii) intensity of cellular response (a1) $x_n$ and $y_n$ and (iv) influence of other environment factors which can interfere with the process of communication. In this case we postulate parameter $r$, that can be taken collectively for intra- and inter- cellular environment, inside of the one variable, indicating overall disposition of the environment to the communication process.

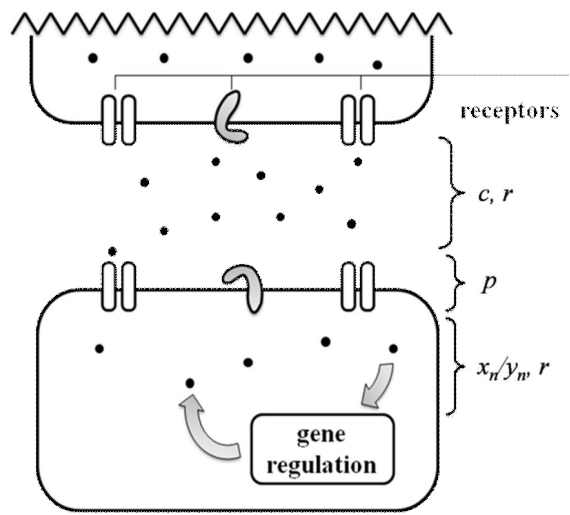

Figure 1. Schematic representation of intercellular communication, taken from [10]. Here, $c$ represents concentration of signaling molecule in intercellular environment coupled with intensity of response they can provoke while $r$ includes collective influence of environment factors which can interfere with the process of communication. $x_n$ and $y_n$ represent concentration of signaling molecules in cells environment, while $p$ denotes cellular affinity to uptake the substances.

The time development ($n$ is the number of time step) of the concentration in cells $(x_n, y_n)$ can be expressed as

$$x_{n+1} = (1-c)\Psi(x_n) + h(\Psi(y_n)), \quad (1a)$$
$$y_{n+1} = (1-c)\Psi(x_n) + h(\Psi(x_n)). \quad (1b)$$

The map, $h$ represents the flow of materials from cell to cell, and $h(x)$ and $h(y)$ are defined by a map that can be approximated by a power map,

$$h(x) \sim cx^p, \quad (2a)$$
$$h(y) \sim cy^q. \quad (2b)$$

If $h(x) \sim cx^p$ and $h(y) \sim cy^q$, the interaction is expressed as a nonlinear coupling between two cells. The dynamics of intracellular behavior is expressed as a logistic map (e.g., [11,12]),

$$\Psi(x_n) = r\, x_n(1 - x_n), \quad (3a)$$
$$\Psi(y_n) = r\, y_n(1 - y_n). \quad (3b)$$

Since concentration of signaling molecules can be regarded as their population for fixed volume, and since we are focused on mutual influence of these populations, it points out to use the coupled logistic equations. Instead of considering cell-to-cell coupling of two explicit n-gene oscillators [13] we consider generalized case of gene oscillators coupling. In that case investigation of conditions under which two equations are synchronized and how this synchronization behaves under changes of intra- and inter- cellular environment, can give some answers on the question of maintaining functionality in the system. Therefore, having in mind that (i) cellular events are discrete [14] and (ii) the aforementioned reasoning, we consider system of difference equations of the form

$$\mathbf{X}_{n+1} = \mathbf{F}(\mathbf{X}_n) \equiv \mathbf{L}(\mathbf{X}_n) + \mathbf{P}(\mathbf{X}_n), \quad (4)$$

with notation

$$\mathbf{L}(\mathbf{X}_n) = ((1-c)rx_n(1-x_n), (1-c)ry_n(1-y_n)), \quad \mathbf{P}(\mathbf{X}_n) = (cy_n^p, cx_n^{1-p}), \quad (5)$$

where $\mathbf{X}_n = (x_n, y_n)$ is a vector representing concentration of signaling molecules inside of the cell, while $\mathbf{P}(\mathbf{X}_n)$ denotes stimulative coupling influence of members of the system which is here restricted only to positive numbers in the interval $(0,1)$. The starting point $\mathbf{X}_0$ is determined so that $(x_0, y_0) \in (0,1)$. Parameter $r \in (0,4)$ is so-called logistic parameter, which in logistic difference equation determines an overall disposition of the environment to the given population of signaling molecules and exchange processes. Affinity to uptake signaling molecules is indicated by $p$. Let us note that we require that sum of all affinities of cells $p_i$ exchanging substances has to satisfy condition $\sum_i p_i = 1$ or in the case of two cells $p + q = 1$. Since fixed point is $\mathbf{F}(0) = 0$, in order to ensure that zero is not at the same time the point of attraction, we defined $p \in (0,1)$ as an exponent. Finally, $c$ represents coupling of two factors: concentration of signaling molecules in intracellular environment and intensity of response they can provoke. This form is taken because the effect of the same intracellular concentration of signaling molecules can vary greatly with variation of affinity of genetic regulators for that signal, which is further reflected on the ability to synchronize with other cells. Therefore, $c$ influence both, rate of intracellular synthesis of signaling molecules, as well as synchronization of signaling processes between two cells, so the parameter $c$ is taken to be a part of both $\mathbf{L}(\mathbf{X}_n)$ and $\mathbf{P}(\mathbf{X}_n)$. However, relative ratio of these two influences depends on current model setting. For example, if for both cells $\mathbf{X}_n$ is strongly influenced by intracellular concentration of signals, while they can provoke relatively smaller response then the form of equation will be

$$x_{n+1} = (1-c)rx_n(1-x_n) + cy_n^p, \quad (6a)$$
$$y_{n+1} = (1-c)ry_n(1-y_n) + cx_n^{1-p}. \quad (6b)$$

We now analyze our coupled system given by (6a)-(6b). For $0 < x < 1$ and $0 < p < 1$ we have $0 < x < x^p < 1$. So, for small $c$, the dynamic of our investigated system is similar to the dynamic of the following systems obtained by minorization, i.e.

$$\begin{aligned} x_{n+1} &= (1-c)rx_n(1-x_n), \\ y_{n+1} &= (1-c)ry_n(1-y_n), \end{aligned} \quad (7)$$

and

$$\begin{aligned} x_{n+1} &= (1-c)rx_n(1-x_n) + cy_n, \\ y_{n+1} &= (1-c)ry_n(1-y_n) + cx_n. \end{aligned} \quad (8)$$

If we apply a majorization the considered system becomes

$$\begin{aligned} x_{n+1} &= (1-c)rx_n(1-x_n) + c, \\ y_{n+1} &= (1-c)ry_n(1-y_n) + c, \end{aligned} \quad (9)$$

For all of those systems is obvious that they do not depend on parameter $p$. Because of $f(x, y) = g(y, x)$, where, $f, g$ are components of **F** in (4), their dynamics are symmetric to the diagonal $\Delta$, $\Delta = \{(x, y): y = x\}$, as it was analyzed in [15]. This symmetry we also have for system (6a)-(6b) if $p = 0.5$.

Having in mind the aforementioned conditions for $p$, $x$ and $x^p$ we consider only systems (7) and (9) since the behavior of the system (6a)-(6b) comes from the properties of the mentioned ones. It is seen that the systems (7) and (9) consist of uncoupled logistic maps, in (7), on the interval $(0, 1)$, while in (9) on the interval $(\delta, 1-\delta)$, where $\delta < 0$ is the smaller solution of the equation $x = (1-c)rx(1-x) + c$, i.e.

$$\delta = [(1-c) - 1 - \sqrt{[r(1-c)-1]^2 + 4cr(1-c)}] / [2r(1-c)]. \quad (10)$$

From the property of the logistic equation we get the following expression

$$\Gamma = [r(1-c) + 4c - 4\delta] / (1 - 2\delta), \quad (11)$$

where $\Gamma$ in system (9) taking the role of $r(1-c)$ in (7). Comparing $r(1-c)$ with (11) we can conclude that the expression (11) is always greater, that implies that bifurcations and chaos first appears for system (7) and than for (9). Finally, combining expressions (10) and (11) we get

$$\Gamma = \frac{r^2(1-c)^2 + 4cr(1-c) - 2r(1-c) + 2 + 2\sqrt{[r(1-c)-1]^2 + 4cr(1-c)}}{1 + \sqrt{[r(1-c)-1]^2 - 4cr(1-c)}} \quad (12)$$

## 3. A NONLINEAR DYNAMICS-BASED ANALYSIS OF THE COUPLED MAPS REPRESENTING THE INTERCELLULAR EXCHANGE OF SUBSTANCES

In order to further investigate the behavior of the coupled maps, we perform a numerical analysis of the coupled system (6) throuhg its parameters $c$, $r$ and $p$, using the largest Lyapunov exponet as measure of the chaotic behaviour and border between synchronized and nonsynchronized system states in intercellular exchange of substances.

Previously [10] we showed that for fixed value of $r$ (in our case 3.95) Lyapunov exponent of the coupled maps, given as a function of the coupling parameter $c$ ranging from 0 to 1.0, for different values of the affinity $p$ gives a border in values of concentration $c$ (around 0.4), that split domain of concentration into two regions. The first one, is located between 0 and 0.4, with the non sinchronyzed states including sporadical windows where synchronization is reached. In contrast to that, the second region (between 0.4 and 1.0) is region where process of ehcange between two cells is fully synchronized.

We calculate Lyapunov exponent by analysis of orbits. The orbit of the point $\mathbf{X}_0$ is the sequence $\mathbf{X}_0, \mathbf{F}(\mathbf{X}_0), ..., \mathbf{F}^n(\mathbf{X}_0), ...$ where $\mathbf{F}^0(\mathbf{X}_0) \equiv \mathbf{X}_0$ and for $n \geq 1$, $\mathbf{F}^n(\mathbf{X}_0) = \mathbf{F}(\mathbf{F}^{n-1}(\mathbf{X}_0))$. We say that the orbit is periodic with period $k$ if $k$ is the smallest natural number such that $\mathbf{F}^k(\mathbf{X}_0) = \mathbf{X}_0$. If $k = 1$, then the point $\mathbf{X}_0$ is the fixed point. The periodic point $\mathbf{X}_0$ with period $k$ is an attraction point if the norm of the Jacobi matrix for the mapping $\mathbf{F}^k(\mathbf{X}) = (f_k(x,y)), (g_k(x,y))$ is less than one, i.e., $\|\mathbf{J}^k(\mathbf{X}_0)\| < 1$, where

$$\mathbf{J}^k(\mathbf{X}_0) = \begin{bmatrix} \dfrac{\partial f_k}{\partial x} & \dfrac{\partial f_k}{\partial y} \\ \dfrac{\partial g_k}{\partial x} & \dfrac{\partial g_k}{\partial y} \end{bmatrix}_{\mathbf{X}=\mathbf{X}_0}. \qquad (13)$$

Here, we define $\|\mathbf{J}^k(\mathbf{X}_0)\|$ as max $\{|\lambda_1|, |\lambda_2|\}$, where $\lambda_1$ and $\lambda_2$ are the eigenvalues of the matrix. It is worth noting that

$$\mathbf{J}^k(\mathbf{X}_k) = \mathbf{J}^k(\mathbf{X}_0) = \mathbf{J}(\mathbf{X}_{k-1})... \mathbf{J}(\mathbf{X}_1)\mathbf{J}(\mathbf{X}_0), \qquad (14)$$

where

$$\mathbf{J}(\mathbf{X}) = \begin{bmatrix} (1-c)r(1-2x) & cpy^{p-1} \\ c(1-p)x^{-p} & (1-c)r(1-2y) \end{bmatrix}. \qquad (15)$$

In particular, for the scalar equation $x_{n+1} = d(x_n)$ the norm is $|(d^k(x))'|_{x=x_0} = |d'(x_{n-1})...d'(x_1)d'(x_0)|$, where $d'(x) = r(1-2x) + cpx^{p-1}$. In order to characterize the asymptotic behavior of the orbits, we need to calculate the largest Lyapunov exponent, which is given for the initial point $\mathbf{X}_0$ in the attracting region by

$$\lambda = \lim_{n \to \infty}(\ln \|\mathbf{J}^n(\mathbf{X}_0)\|/n). \qquad (16)$$

With this exponent, we measure how rapidly two nearby orbits in an attracting region converge or diverge. In practice, using (8), we compute the approximate value of $\lambda$ by substituting in (16) successive values from $\mathbf{X}_{n_0}$ to $\mathbf{X}_{n_1}$, for $n_0, n_1$ large enough to eliminate transient behaviors and provide good approximation. If $\mathbf{X}_0$ is part of a stable periodic orbit of period $k$, then $\|\mathbf{J}^k(\mathbf{X}_0)\| < 1$ and the exponent $\lambda$ is negative, which characterizes the rate at which small perturbations from the fixed cycle decay, and we can call such a system synchronized one.

Since robustness, according to the aforementioned is a property that allows a system to maintain its functions against internal and external perturbations we investigate the synchronization as an indicator of robustness of the coupled maps (6) to perturbation of $c$, $r$ and $p$ parameters. To do that, we calculate Lyapunov exponent as a function of the (i) coupling parameter $c$ ranging from 0.0 to 1.0 and (ii) parameter $r$ ranging from 3 to 4, for value of the affinity $p = 0.5$. The changes in the Lyapunov exponent of the coupled maps (6), as a function of parameters $c$ and $r$ is given in Figure 2. From this figure is seen that the model, describing intercellular exchange of substances, does not maintain its functions for lower concentrations $c$ and higher values of the parameter $r$. Figure 3 shows Lyapunov exponent of the coupled maps (6) for different values of the affinity $p = 0.4, 0.3, 0.2, 0.1$.

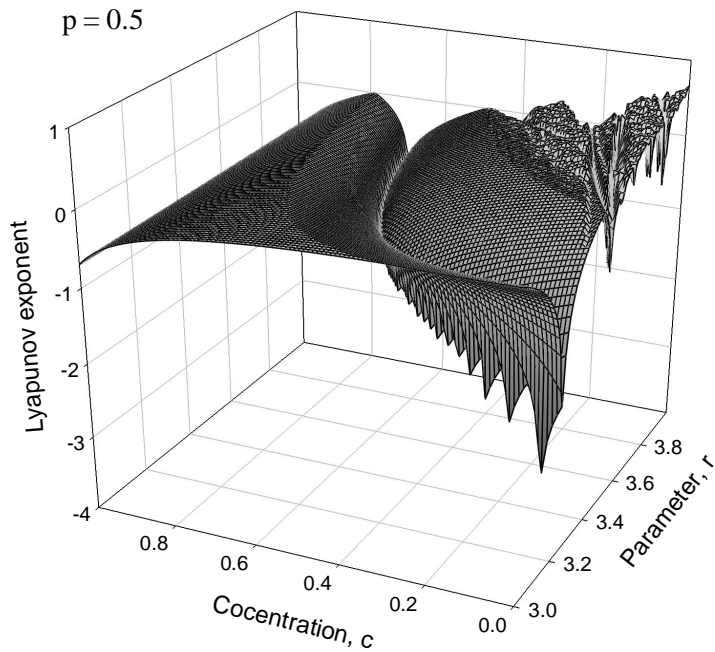

Figure 2. Lyapunov exponent of the coupled maps (6), given as a function of the (i) coupling parameter *c* ranging from 0.0 to 1.0 and (ii) parameter *r* ranging from 3 to 4, for value of the affinity

$p = 0.5$. Each point in the above graphs was obtained by iterating many times (2000 iterations) from the initial condition to eliminate transient behavior and then averaging over another 500 iterations. Initial condition: $x = 0.3$, $y = 0.5$, with 200$c$ values.

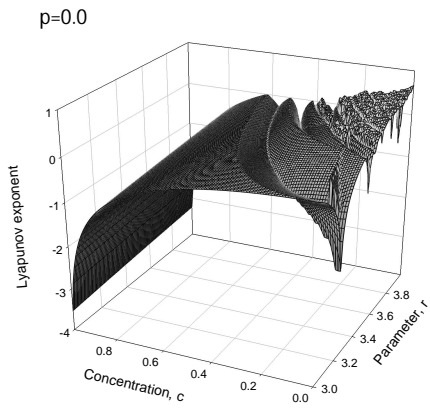

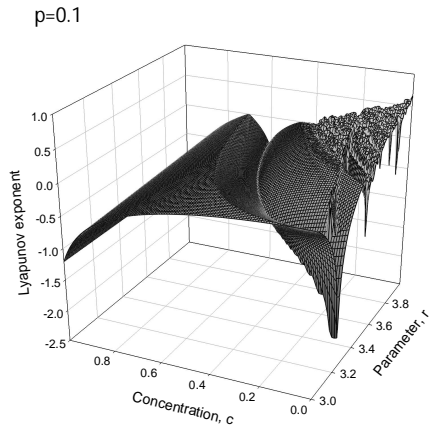

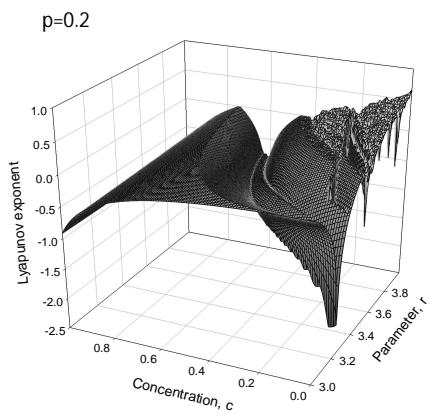

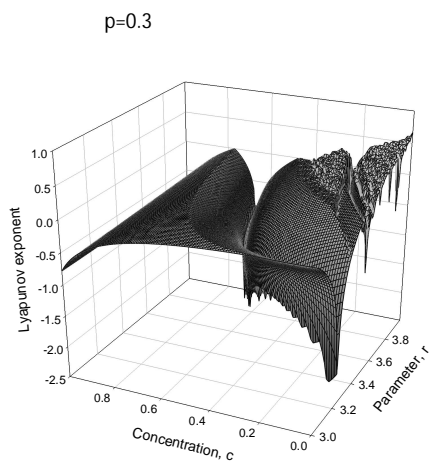

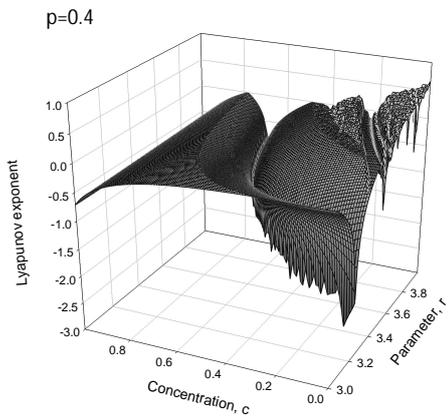

Figure 3. Lyapunov exponent of the coupled maps, given as a function of the (i) coupling parameter $c$ ranging from 0.0 to 1.0 and (ii) parameter $r$ ranging from 3 to 4, for different values of the affinity $p$. The same graphs will be able to obtained for values $p = 0.6, 0.7, 0.8, 0.9$ and $1.0$ corresponding to those for $p = 0.4, 0.3, 0.2, 0.1$ and $0.0$. Each point in the above graphs was obtained by iterating many times (2000) from the initial condition to eliminate transient behavior and then averaging over another 500 iterations. Initial condition: $x = 0.3$, $y = 0.5$, with $200c$ values.

## 4. CONCLUSIONS

In this paper, our focus is on investigating stability of intercelular exchange of biochemical substances affected by variability of environmental parameters. We identified main parameters of the process of cellular communication and using a system of two coupled logistic equations we investigated synchronization of the model and its sensitivity to fluctuations of environmental parameters. Results show existence of stability regions where noise in the form of fluctuations in concentration of signaling molecules in intercellular environment and fluctuations in affinity for uptake these molecules cannot interfere with the process of exchange. Since our model is insipred by the general scheme of intercellular communication, it naturally does not allow detailed modelling of some concrete, emiprically verifable intercellular communication process. Instead, it is designed to serve as a starting tool in general investigation of robustness in mutually stimulative populations which can be readily extended to investigation of synchronization in larger networks of interacting entities [16,17].

**Acknowledgments.** The research work described here has been funded by the Serbian Ministry of Science and Technology under the project "Numerical simulation and monitoring of climate and environment in Serbia", for 2011-2014.